\documentclass[%
 reprint,
superscriptaddress,
 amsmath,amssymb,
 aps,
 prl,
]{revtex4-2}

\usepackage{graphicx,color}
\usepackage{dcolumn}
\usepackage{bm}
\usepackage{times}

\begin{document}

\preprint{APS/123-QED}

\title{Many-body correlations are non-negligible in both fragile and strong glassformers}

\author{Chengjie~Luo}
\email[Electronic mail: ]{C.Luo@tue.nl}
\affiliation{Soft Matter and Biological Physics, Department of Applied Physics, Eindhoven University of Technology, P.O. Box 513, 5600 MB Eindhoven, The Netherlands}
\author{Joshua~F.~Robinson}
\affiliation{Institut f\"ur Physik, Johannes Gutenberg-Universit\"at Mainz, Staudingerweg 7-9, 55128 Mainz, Germany}
\affiliation{H. H. Wills Physics Laboratory, University of Bristol, Bristol BS8 1TL, United Kingdom}
\author{Ilian~Pihlajamaa}
\affiliation{Soft Matter and Biological Physics, Department of Applied Physics, Eindhoven University of Technology, P.O. Box 513, 5600 MB Eindhoven, The Netherlands}
\author{Vincent~E.~Debets}
\affiliation{Soft Matter and Biological Physics, Department of Applied Physics, Eindhoven University of Technology, P.O. Box 513, 5600 MB Eindhoven, The Netherlands}
\author{C.~Patrick~Royall}
\affiliation{Gulliver UMR CNRS 7083, ESPCI Paris, Universit\'{e} PSL, 75005 Paris, France}
\affiliation{H. H. Wills Physics Laboratory, University of Bristol, Bristol BS8 1TL, United Kingdom}
\affiliation{School of Chemistry, Cantock's Close, University of Bristol, Bristol, BS8 1TS, United Kingdom}
\affiliation{Centre for Nanoscience and Quantum Information, University of Bristol, Bristol BS8 1FD, United Kingdom}
\author{Liesbeth M.~C.~Janssen}
\email[Electronic mail: ]{L.M.C.Janssen@tue.nl}
\affiliation{Soft Matter and Biological Physics, Department of Applied Physics, Eindhoven University of Technology, P.O. Box 513, 5600 MB Eindhoven, The Netherlands}

\date{\today}

\begin{abstract}
It is widely believed that the emergence of slow glassy dynamics is encoded in a material’s microstructure. 
First-principles theory [mode-coupling theory (MCT)] is able to predict the dramatic slowdown of the dynamics from only static two-point correlations as input, yet it cannot capture all of the observed dynamical behavior.
Here we go beyond two-point spatial correlation functions by extending MCT systematically to include higher-order static and dynamic correlations. We demonstrate that only adding the static triplet direct correlations already qualitatively changes the predicted glass-transition diagram of binary hard spheres and silica. 
Moreover, we find a non-trivial competition between static triplet correlations that work to \emph{stabilize} the glass state, and dynamic higher-order correlations which \emph{destabilize} it for both materials. We conclude that the conventionally neglected static triplet direct correlations as well as higher-order dynamic correlations are in fact non-negligible in both fragile and strong glassformers.

\end{abstract}

\maketitle

\textit{Introduction.---} Almost any material can be supercooled or compressed from the liquid to the glass state. During this process the relaxation dynamics slows down dramatically, while the disordered microstructure undergoes only small changes. This apparent disconnect between structure and dynamics underlies much of the complexity of the glass transition \cite{binder2011glassy,debenedetti2001supercooled,berthier2011theoretical}.

Among the key challenges in tackling the glass transition is the dearth of first--principles theories. Arguably the dominant first--principles theory, mode--coupling theory (MCT) \cite{bengtzelius1984dynamics,leutheusser1984dynamical,gotze2008complex,reichman2005mode,janssen2018mode}, 
%in its standard form \paddycomment{[]Paddy $<---$ there is for sure a better way of putting this!]}, 
does a good job of predicting the increase in structural relaxation time for the first 4--5 decades and indeed can be accurately fitted to experimental \cite{vanmegen1998} and simulation \cite{kob1995testing} data of \textit{fragile} glassformers in this regime. However, at deeper supercooling, the power--law increase in relaxation time predicted by MCT leads to a total dynamical arrest at state points where experiments and computer simulations still exhibit relaxation
\cite{brambilla2009probing, hallett2018, ortlieb2021relaxation}. Moreover, this power law is completely incompatible with the Arrhenius behavior of  \textit{strong} glassformers \cite{angell1988structural}, even at relatively high temperatures. MCT nonetheless predicts qualitatively reliable state diagrams for both fragile and strong glassformers \cite{gotze2003effect,voigtmann2008dynamics}.

The static structure factor (i.e.\ two-point density correlations) is normally
assumed to be a sufficient representation of the structure to describe a system with pairwise interactions \cite{hansen1990theory}, as used in the usual implementation of MCT.
However, it has been shown both experimentally and numerically \cite{leocmach2012,hallett2018,coslovich2007,zhang2020revealing} that upon supercooling, higher--order structural motifs change markedly (much more than two--point correlations). Moreover, machine--learning methods identify subtle structural changes \cite{schoenholz2016structural,bapst2020unveiling,paret2020assessing,boattini2020autonomously} and configurational entropy drops approaching dynamical arrest \cite{banerjee2014role,berthier2017configurational,williams2018experimental}. Accompanying these subtle changes of structure, it is recognized that higher--order dynamical correlations implicit in so--called cooperatively rearranging regions may
enable relaxation at supercoolings past the critical volume fraction $\varphi_\mathrm{mct}$ (or critical temperature $T_\mathrm{mct}$) at which MCT predicts the dynamical divergence
\cite{berthier2011theoretical,royall2015role}.
Given the clear failure of standard MCT to predict dynamical behaviour at deep supercooling and these observations of higher--order correlations, there is a clear need to develop a first--principles theory which captures such correlations in a consistent and systematic manner.

The first such attempt to go beyond the two-body level within MCT is to consider static triplet correlations $c^{(3)}$, with  $c^{(3)}$ calculated either from simulations \cite{coslovich2013static,sciortino2001debye} or theories such as density functional theory \cite{jorge2002theory,hansen1990theory}. In the 1980s, Barrat \textit{et al}.\  \cite{barrat1989liquid} concluded that $\varphi_\mathrm{mct}$ for Percus-Yevick 
hard spheres only quantitatively changes from $0.516$ to $0.512$ by supplementing MCT with $c^{(3)}$. 
Later, Sciortino and Kob \cite{sciortino2001debye} found that for both the BKS model of silica and the Kob-Andersen binary Lennard-Jones model the inclusion of triplet correlations improves the prediction of the non-ergodicity parameters; the improvement is particularly significant for the former, but at a $T_{\mathrm{mct}}$ that is further removed from the simulated glass transition temperature.
Recently, Ayadim \textit{et al}.\ \cite{ayadim2011mode} discovered that $c^{(3)}$ also quantitatively affects the location of the MCT glass-transition line and the non-ergodicity parameter for hard-core particles with short-ranged interactions. These findings hint that many-body correlations such as $c^{(3)}$ could be important for the glassy dynamics of even simple glassformers, but  the precise importance and role of such correlations remains ambiguous.

Recently, generalized mode-coupling theory (GMCT) has been developed as a systematic extension of MCT that adds in higher-order dynamic and static correlations \cite{szamel2003colloidal,wu2005high,janssen2015microscopic,ciarella2021multi,debets2021generalized}. Under certain conditions, the hierarchical GMCT framework is even able to account for relaxation behaviors other than the unrealistic power law noted above \cite{mayer2006cooperativity}.  
So far, within GMCT, only multi-point \emph{dynamic} correlation functions have been
incorporated, leading to quantitatively improved critical point predictions with respect to MCT for fragile glassformers. This suggests an improved ability of GMCT to amplify small differences in static structure factors
\cite{szamel2003colloidal,wu2005high,janssen2015microscopic,luo2020generalized1,luo2020generalized2, luo2021glassy,ciarella2021multi,berthier2010critical}.
However, it remains unclear how the combination of \textit{both} structural and dynamic higher--order correlations affects the glassy dynamics from a first--principles perspective. 

Here we make a first step towards a full many--body, first--principles treatment of glassy systems by including both static three--body terms and dynamic many-body terms into the GMCT framework. Our method is quite general and can readily be extended to include static four--body and still higher--order terms \cite{rosenfeld1989free}.  We consider two model glassformers from different fragility classes; one is a binary hard sphere mixture (BHS) with a range of size ratios $0.5\leq\delta\leq0.8$, 
and the other is the strong BKS model of silica \cite{van1990force}.
We find that $c^{(3)}$ has a non-negligible effect, both quantitatively and qualitatively, on the prediction of the liquid-glass transition even for BHS with small size disparities. Furthermore, the GMCT level of the multi-point \textit{dynamic} correlation functions also qualitatively changes the diagrams of BHS and significantly improves the prediction of relaxation times for $\text{SiO}_2$, indicating that both static and dynamic multi-point density correlations play a significant and non-trivial role in both glassformers. An implication of our work is that the reasonable predictions of MCT for state diagrams and non-ergodicity parameters \cite{gotze2008complex}, based solely on two-point correlations, could essentially be regarded as a coincidence or a cancellation of errors. 

\textit{First-principles theory.---}We first briefly introduce the GMCT framework. The central dynamic objects for a multi-component glassy system consisting of $\mathcal{N}_p$ particles and $M$ species are the multi-component $2n$-point density correlation functions
$    {F}^{(2n)}_{\{\alpha_i\};\{\beta_i\}}(\{k_i\},t) =\left<\rho^{\alpha_1}_{-\bm{k_1}}(0)\rho^{\alpha_2}_{-\bm{k_2}}(0)\hdots \rho^{\alpha_n}_{-\bm{k_n}}(0)\rho^{\beta_1}_{\bm{k_1}}(t)\rho^{\beta_2}_{\bm{k_2}}(t)\hdots \rho^{\beta_n}_{\bm{k_n}}(t)\right>$, 
where 
$ \rho^{\alpha}_{\bm{k}}(t)=\sum_{{i_\alpha}=1} ^{\mathcal{N}_{\alpha}}e^{i\bm{k}\cdot\bm{r}_{i_\alpha}(t)}/\sqrt{\mathcal{N}_p}$
 is a density mode for species $\alpha \in \{1,2,\hdots,M\}$ at wavevector $\bm{k}$ and time $t$, the angular brackets denote an ensemble average, $\bm{r}_{i_\alpha}(t)$ is the position of particle $i_\alpha$ of species $\alpha$ at time $t$, and $\mathcal{N}_\alpha$ is the number of particles of type $\alpha$ with $\sum_{\alpha=1}^{M}\mathcal{N}_{\alpha}=\mathcal{N}_p$. The label $n \in \{1,2,\hdots,\infty\}$ is the level of the GMCT hierarchy; when $n=1$,  $F^{(2)}_{\alpha;\beta}(k,t)$ is the
 partial intermediate scattering function. 
These dynamic equations have been derived from repeated application of the Mori-Zwanzig approach, where we have neglected  
the off-diagonal correlations \cite{ciarella2021multi,janssen2015microscopic}.

 The hierarchical equations for ${F}^{(2n)}_{\{\alpha_i\};\{\beta_i\}}(\{k_i\}, t)$  read \cite{ciarella2021multi}
\begin{align}
    &\ddot{F}^{(2n)}_{\{\alpha_i\};\{\beta_i\}}(t)+F^{(2n)}_{\{\alpha_i\};\{\gamma_i\}}(t)\left(S^{(2n)}\right)^{-1}_{\{\gamma_i\};\{\theta_i\}}J^{(2n)}_{\{\theta_i\};\{\beta_i\}}\nonumber\\
&+\int_0^t d\tau \dot{F}^{(2n)}_{\{\alpha_i\};\{\gamma_i\}}(t-\tau) \left(J^{(2n)}\right)^{-1}_{\{\gamma_i\};\{\theta_i\}}K^{(2n)}_{\{\theta_i\};\{\beta_i\}}(\tau)=0,
\label{eq:Flong}
\end{align}
where the arguments $\{k_i\}$ and the summation over indices of species are omitted for simplicity.  $S^{(2n)}_{\{\alpha_i\};\{\beta_i\}}(\{k_i\})
\equiv F^{(2n)}_{\{\alpha_i\};\{\beta_i\}}(\{k_i\},t=0)$
is the $2n$-point static density correlation function describing the microstructure of the system and $J^{(2n)}_{\{\alpha_i\};\{\beta_i\}}(\{k_i\})=
    \left<\frac{d}{dt}\left[\rho^{\alpha_1}_{-\bm{k}_1}...\rho^{\alpha_n}_{-\bm{k}_n}\right]\frac{d}{dt}\left[\rho^{\beta_1}_{\bm{k}_1}...\rho^{\beta_n}_{\bm{k}_n}\right]\right>$ is the general static current-current matrix. Both of them contain only even-order density modes which can be approximated using Gaussian factorization, and hence they only depend on the two-point static density correlations $S^{(2)}_{\alpha;\beta}(k)$ \cite{ciarella2021multi}. 
    The key unknown part of Eq.~(\ref{eq:Flong}) is the memory kernel $K^{(2n)}$, which GMCT hierarchically expands as a linear combination of the next-level correlators $F^{\left(2(n+1)\right)}$ \cite{szamel2003colloidal,janssen2015microscopic,ciarella2021multi,debets2021generalized}. That is,
\begin{align}
&K^{(2n)}_{\{\alpha_i\};\{\beta_i\}}(\{k_i\},t)=\frac{\rho }{16\pi^3}
\sum_{\substack{
    \mu\nu\\
    \mu'\nu'}
}
\sum_{j=1}^{n}\int d\bm{q}
\nonumber \\
&\quad\times\frac{k_BT}{m_{\alpha_j}}\mathcal{V}_{\mu'\nu'\alpha_j}({\bm{q},\bm{k_j}-\bm{q}})\mathcal{V}_{\mu\nu\beta_j}({\bm{q},\bm{k_j}-\bm{q}})\frac{k_BT}{m_{\beta_j}}
\nonumber \\
&\quad\times F^{(2(n+1))}_{\mu',\nu',\{\alpha_i\}/\alpha_j;\mu,\nu,\{\beta_i\}/\beta_j}({\bm{q},\bm{k_j}-\bm{q}},\{k_i\}/k_j,t) ,
\label{eq:kn}
\end{align}
where $T$ is the temperature and $m_\alpha$ is the mass of particle species $\alpha$. The subscript $\{x_i\}/x_j$ is a list $x_1,x_2,\hdots,x_n$ with the specific element $x_j$ removed.

\begin{figure*}[hbtp]
	\includegraphics[width=0.9\textwidth]{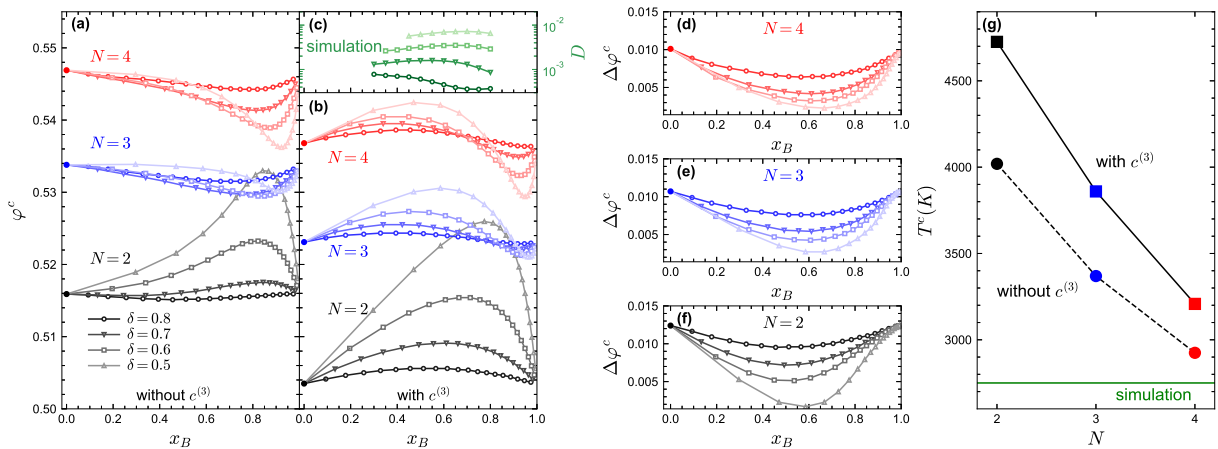}
	\caption{\label{fig:phic_xB_withD} 
	Liquid-to-glass state diagrams of binary hard spheres as predicted from GMCT (a) without $c^{(3)}$ and (b) with $c^{(3)}$. The curves represent the critical packing fractions $\varphi^c$ as a function of the number concentration of the small species, $x_B=\mathcal{N}_B/\mathcal{N}_p$. 
	Different symbols correspond to different particle size ratios $\delta$, and different colors correspond to different GMCT closure levels $N$.
	(c) The long-time diffusion coefficient $D$ from simulations at packing fraction $\varphi=0.57$. (d)-(f) The difference of the critical packing fractions without and with  $c^{(3)}$, i.e., $\Delta \varphi^c(N,\delta,{x}_B)=\varphi^c(N,\delta,{x}_B,\text{without  } c^{(3)})-\varphi^c(N,\delta,{x}_B,\text{with } c^{(3)})$, as a function of ${x}_B$. The filled circles indicate the results for one-component hard spheres.  (g) Critical temperature $T^c$ of $\text{SiO}_2$ predicted from GMCT as a function of level $N$. Squares and circles are with and without $c^{(3)}$, respectively. The solid green line indicates the lowest temperature $T=2750K$ at which the system can be equilibrated in simulation, and hence it is an upper bound of the glass transition temperature.
	}
\end{figure*}
The static vertices, which represent the coupling strength between different wavevectors, are given by \cite{ciarella2021multi,sciortino2001debye}
\begin{align}
&\mathcal{V}_{\alpha\beta\mu}(\bm{q},\bm{k-q})=
{\rho}k^2x_{\mu}c^{(3)}_{\alpha\beta\mu}(\bm{q},\bm{k-q})
\nonumber\\&
+(\bm{k-q})\cdot\bm{k}c^{(2)}_{\beta\mu}({|\bm{k-q}|})\delta_{\alpha\mu}
+ \bm{q}\cdot\bm{k}c^{(2)}_{\alpha\mu}(q)\delta_{\beta\mu},
\label{eq:vertex}
\end{align}
where $x_\mu=\mathcal{N}_\mu/\mathcal{N}_p$ is the number concentration of species $\mu$, $c^{(2)}_{\alpha\beta}(q)$ is the (doublet) direct correlation function connecting to $S^{(2)}$ via the Ornstein-Zernike equation \cite{barrat1988equilibrium,hansen1990theory} 
and 
$c^{(3)}_{\alpha\beta\mu}(\bm{q},\bm{k-q})$
is the triplet direct correlation function \cite{barrat1988equilibrium,sciortino2001debye}. 

It is clear now that the inputs to the GMCT framework are only $S^{(2)}$ and $c^{(3)}$, with the latter only appearing in the vertices. 
To numerically solve the GMCT equations, we follow previous work \cite{ciarella2021multi,janssen2015microscopic,debets2021generalized} and apply a self-consistent closure for the multi-point dynamic density correlations at the highest level $N$, such that $F^{(2N)}(t)\approx F^{(2)}(t)\times F^{(2(N-1))}(t)$ (see Supplementary Information (SI) \footnote{See Supplemental Material [url] for the details on the theories, simulations and results of this work which includes references \cite{biezemans2020glassy,janssen2014relaxation,janssen2016generalized,flenner2005relaxation,franosch1997asymptotic,lebowitz1964exact,foffi2004alpha,voigtmann2011multiple,thompson2022lammps}}). In the following we report how the inclusion of $c^{(3)}$ and the level $N$ affect the liquid-glass transition for BHS and $\text{SiO}_2$, and the predicted fragility of the latter. As inputs, we take $S^{(2)}$ and $c^{(3)}$ from  Rosenfeld's fundamental measure theory \cite{rosenfeld1989free,rosenfeld1990free} for BHS and from molecular dynamics simulations for $\text{SiO}_2$ (see SI for details).

\textit{Glass-transition diagrams.---} 
We first focus on BHS to discuss the role of the GMCT closure level $N$. Figures \ref{fig:phic_xB_withD}(a) and \ref{fig:phic_xB_withD}(b) show the GMCT-predicted liquid-glass transition diagrams for several values of the particle size ratio $\delta=d_B/d_A$, where $d_A$ and $d_B$ denote the particle diameters of the larger and smaller species respectively.
Increasing $N$ generally leads to higher critical packing fractions, both with and without $c^{(3)}$, which is consistent
with GMCT predictions for monodisperse hard spheres \cite{szamel2003colloidal,wu2005high,luo2020generalized1} and sticky hard spheres \cite{luo2021glassy}, and also in better quantitative agreement with the critical points $\varphi^c>0.58$ reported in experiments and simulations \cite{foffi2003mixing,williams2001motions,henderson1998metastability}. 
Therefore, regardless of $c^{(3)}$, including higher level \textit{dynamic} density correlations effectively stabilizes the \textit{liquid} state.

Let us check more carefully the effect of $N$ for different size ratios. In standard MCT, i.e.\ for $N=2$ and neglecting $c^{(3)}$, two opposite effects are observed at different size disparities \cite{voigtmann2003mode,gotze2003effect}. Specifically, at a small size disparity such as $\delta=0.8$ (open circles in  Fig.~\ref{fig:phic_xB_withD}(a)), 
the critical packing fraction is slightly smaller than the monodisperse case (filled circle at $x_B=0$). However, for a large size disparity such as $\delta=0.5$ (upward-pointing triangles in Fig.~\ref{fig:phic_xB_withD}(a)), the mixing of the two species leads to higher critical packing fractions. This latter effect is known as entropically-induced plasticization and can be attributed to the depletion attraction \cite{williams2001motions,voigtmann2003mode}. 
By increasing $N$ and neglecting $c^{(3)}$, we see that the plasticization effect completely disappears and all curves in fact show an inverse-plasticization trend. Contrasting this with the results for small size disparities ($\delta=0.8$), where the inverse plasticization does not change much as $N$ increases, we can conclude that the effect of $N$ depends on the size ratio and becomes more pronounced for large size disparities.  

Interestingly, when we include $c^{(3)}$, the inverse-plasticization effect for $N=3$ and $N=4$ without $c^{(3)}$ is inverted again for $x_B\lesssim 0.7$, as can be seen in Fig.~\ref{fig:phic_xB_withD}(b). Notice that even for MCT ($N=2$), adding $c^{(3)}$ qualitatively changes the transition curve at small size disparities such as $\delta=0.8$. Therefore the triplet direct correlation functions $c^{(3)}$ greatly affect the state diagrams in a non-trivial way. 
 
To delineate the specific effect of $c^{(3)}$ on the glass-transition curves, we plot the difference of the critical packing fractions without and with $c^{(3)}$, $\Delta \varphi^c=\varphi^c(\mathrm{without }\ c^{(3)})-\varphi^c(\mathrm{with }\ c^{(3)})$, as shown in panels (d)--(f) of Fig.~\ref{fig:phic_xB_withD}. All the differences are positive, which means that, contrary to the role of increasing $N$, adding $c^{(3)}$ effectively stabilizes the \textit{glass} state. More surprisingly, both the shapes and the magnitudes of the  $\Delta \varphi^c$ curves as a function of ${x}_B$ are similar for all levels $N$. This may be attributed to the fact that the shapes of the vertices, which contain $c^{(3)}$, remain qualitatively similar for different $\varphi$ and $N$ (see SI).
Moreover, for each level $N$, the largest deviations from the single-component result ($\Delta \varphi^c({x}_B=0)$) are found for the smaller size ratios $\delta$. 
These results indicate that $c^{(3)}$ plays a larger role with increasing size disparities, similar to the greater effect of level $N$ for large size disparities we discussed before.

To judge whether GMCT gives reasonably good predictions for BHS, we perform event-driven molecular dynamics simulations at high packing fractions \cite{bannerman2011dynamo,marin2020tetrahedrality}. Since the shapes of the long-time diffusion coefficient $D$ curves as a function of $x_B$ do not qualitatively change when $\varphi\geq0.57$ \cite{marin2020tetrahedrality}, we can regard the diffusion constants at $\varphi=0.57$ as a semi-quantitative indicator for the glass transition lines, as shown in Fig.~\ref{fig:phic_xB_withD}(c).
Qualitatively, MCT without $c^{(3)}$ seems to perform best because both the plasticization effect at small size ratios ($0.5\leq \delta\leq 0.7$) and the inverse-plasticization effect at large size ratios ($\delta=0.8$) are  observed in the simulations \cite{foffi2003mixing}, whereas for larger $N$ these two effects are not captured simultaneously. 
However, if we look more carefully at the location of the peak of $D$ in Fig.~\ref{fig:phic_xB_withD}(c), we can see that as $\delta$ is increased from $0.6$ to $0.8$, 
the location of the peak decreases from $\sim0.7$ to $\sim0.5$.
Without $c^{(3)}$ for $N=2$ the location of the extremum is not seen to change significantly, whereas incorporating $c^{(3)}$ for $N=3$ introduces more variation in line with the simulations.
Hence GMCT performs better than MCT at least for certain size ratios. We expect that the good predictions from standard MCT are coincidental, perhaps caused by a fortunate cancellation of errors.
We also recall that in the current GMCT framework the multi-point (even-order) static density correlation functions are approximated using Gaussian factorization, the off-diagonal dynamic correlators are ignored \cite{janssen2015microscopic,ciarella2021multi}, and the $S^{(2)}$ and $c^{(3)}$ that we use here might be slightly inaccurate compared to simulations \cite{jorge2002theory,rosenfeld1989free,rosenfeld1990free}. All of these aspects will affect the accuracy of the state diagrams. 

Now let us look at the critical temperature $T^c$ of the strong glassformer $\text{SiO}_2$ predicted from GMCT. From Fig.~\ref{fig:phic_xB_withD}(g) we can see that both with or without $c^{(3)}$, increasing the level $N$ leads to lower critical temperatures, approaching the glass transition point. We point out that the $T^c$ at $N=4$ is even lower than the critical temperature $T_{\text{fit}}=3330 K$ obtained from the power-law fit of diffusion coefficients or $\alpha$-relaxation times \cite{horbach1999static,horbach2001relaxation}, unambiguously showing that GMCT is indeed able to go beyond the MCT regime. However, for a given level $N$, the critical temperature becomes higher when $c^{(3)}$ is included. Note that the higher temperature of silica corresponds to the lower packing fraction of hard spheres, hence the two competitive effects that including $c^{(3)}$ stabilizes the glass state while increasing $N$ stabilizes the liquid state are universal.

\begin{figure}[h!]
	\includegraphics[width=3.375in]{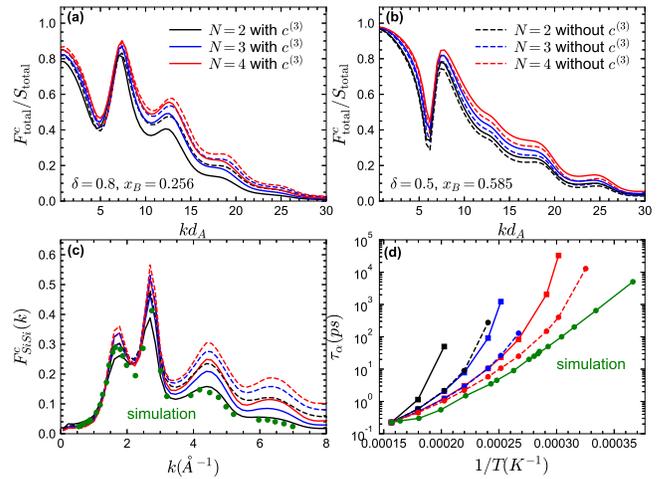}
	\caption{\label{fig:fc_fourfigs}Critical non-ergodicity parameters predicted from GMCT for different closure levels $N$. The top panels show the normalized total NEP $F^c_{\mathrm{total}}/S_{\mathrm{total}}$ for BHS where $F^c_{\mathrm{total}}=\sum_{\alpha\beta}F^{(2)c}_{\alpha;\beta}$ and $S_{\mathrm{total}}=\sum_{\alpha\beta}S^{(2)}_{\alpha;\beta}$ as a function of the wavenumber $k$ for (a) $\delta=0.8$, $x_B=0.256$ and (b) $\delta=0.5$, $x_B=0.585$. 
	In both cases $x_B$ is chosen such that the packing contribution of the small species is $\hat{x}_B=\varphi_B/\varphi=0.15$.
	Panel (c) shows the Si-Si partial NEP $F^c_{SiSi}(k)$ for $\text{SiO}_2$ at the critical points shown in Fig.\ref{fig:phic_xB_withD}(g) predicted from GMCT. Green circles are simulation data from \cite{sciortino2001debye}. (d) The $\alpha$-relaxation time $\tau_\alpha$ as a function of inverse temperature. $\tau_\alpha$ is determined from the partially intermediate scattering function $F_{SiSi}(k=1.771\AA^{-1},\tau_\alpha)/S_{SiSi}(k=1.771\AA^{-1})=e^{-1}$. For better comparison, we scaled the predicted relaxation times for each level $N$ with/without $c^{(3)}$ to make them coincide with the simulated relaxation time at the highest temperature.}
\end{figure}

\textit{Non-ergodicity parameters and fragility.---}To further study the roles of the dynamics-related $N$ and the statics-related $c^{(3)}$, we consider the non-ergodicity parameters (NEPs) at the critical points. Again, we first focus on BHS. In Fig.~\ref{fig:fc_fourfigs}(a) and (b) the normalized total NEPs are shown for two different size ratios, $\delta=0.8$ and $\delta=0.5$, but at the same packing contribution of the smaller species, i.e., $\hat{x}_B=\varphi_B/\varphi=0.15$. It can be seen that for both cases, and both with or without $c^{(3)}$, the NEPs increase as $N$ increases, concurrent with the $N$-dependent increase in the critical packing fraction.

The effect of $c^{(3)}$ on the NEP is, however, more complex. For $\delta=0.8$,  similar to the monodisperse case (see Fig.~S4), adding  $c^{(3)}$ decreases the NEP and hence the relaxation dynamics becomes relatively faster. However, for a small size ratio of $\delta=0.5$, the opposite effect of $c^{(3)}$ can be observed: when $c^{(3)}$ is included in the vertices, the NEP increases for a given level $N$. Note that the critical packing fraction obtained with $c^{(3)}$ is \textit{always} lower than the one without $c^{(3)}$, hence the effect of $c^{(3)}$ on the NEP is not exactly the same as that on the critical packing fractions. In fact, for a large size dispersity of $\delta=0.5$, the effect of $c^{(3)}$ becomes highly number-concentration dependent (see Fig.~S5). Therefore, consistent with our previous observation that $c^{(3)}$ plays an increasingly important role in the glass-transition diagrams at larger size disparities, we find here that a larger size disparity even produces different effects of $c^{(3)}$ on the NEP. This further confirms that $c^{(3)}$ is non-negligible especially for large size disparities.

Next we consider the NEP for silica. 
It is clear from Fig.~\ref{fig:fc_fourfigs}(c) that the effect of $c^{(3)}$ and $N$ on the partial NEP is similar to that on the BHS at a small size disparity $\delta=0.8$. 
This is reasonable because the effective size ratio of oxygen and
silicon is around 0.84 if we estimate their effective sizes
using the first peak position of the corresponding partial radial
distribution function \cite{horbach1999static}. 
We also note that for a given level $N$, adding $c^{(3)}$ always make the NEP closer to the simulation results, which further indicates the important role of $c^{(3)}$. 

To illustrate the important role of higher-order dynamic correlations for the glassy dynamics, we finally plot the relaxation times $\tau_\alpha$ as a function of inverse temperature, which reflects the fragility of the material, as shown in Fig.~\ref{fig:fc_fourfigs}(d). 
Contrary to the NEP, the relaxation times predicted from MCT including $c^{(3)}$ 
deviate most from simulation. 
The predicted curves substantially come closer to the simulation data when $N$ increases from 2 to 4. Although in principle for any finite mean-field closure level $N$ we can only obtain a power law of the relaxation time near the critical point \cite{luo2020generalized2}, here we see that GMCT with increasing $N$ is able to improve the shape of the curves over a larger temperature window, i.e., provides a better prediction of the fragility.

\textit{Conclusions.---}In this work, we have made the first step to incorporate both higher-order spatial and temporal correlations into a coherent first-principles framework for the glass transition. Our theory is able to pick up the small structural changes encoded in static multi-point correlation functions to generally improve predictions for the glass transition.

We have demonstrated that static triplet correlations and dynamic multi-point correlations greatly affect the glass transition in a competitive manner for both fragile and strong glassformers. 
 It is well-known that static triplet correlations are important for strong glassformers while high-order dynamic correlations are vital for improving the dynamics of fragile glassformers, but we have shown that the reverse is also true, and hence both static and dynamic many-body correlations are non-negligible in both fragile and strong glassformers. We mention that the many-body correlations in our theory, which are defined in Fourier space, in principle contain all information for the corresponding many-body quantities in real space. Hence, one could formally establish a relation with e.g.\ locally preferred structures \cite{leocmach2012,hallett2018,coslovich2007,zhang2020revealing} via a Fourier transform. Our framework should therefore also be sensitive to small structural changes in real space \cite{schoenholz2016structural,bapst2020unveiling,paret2020assessing,boattini2020autonomously}, but a full real-space analysis is left for future work. 

We hypothesize that using additional static higher-order correlations
and solving the GMCT equations up to higher closure levels $N$ should bring the predicted glass transition point and relaxation dynamics
closer to the empirical data, 
but full convergence is currently hampered by the high computational cost associated with such calculations.

%%% test if miss any references
%\cite{ciarella2021multi,sciortino2001debye,barrat1988equilibrium,hansen1990theory,debets2021generalized,mayer2006cooperativity,biezemans2020glassy,janssen2014relaxation,janssen2016generalized,voigtmann2003mode,flenner2005relaxation,franosch1997asymptotic,rosenfeld1989free,rosenfeld1990free,lebowitz1964exact,bannerman2011dynamo,foffi2004alpha,jorge2002theory,voigtmann2011multiple,gotze2003effect,luo2020generalized1,van1990force,thompson2022lammps,horbach1999static,wu2005high}

\textit{Acknowledgments.---}We thank Gilles Tarjus for insightful discussions.
This work has been financially supported by the Dutch Research Council (NWO) through a START-UP grant (CL, VED, and LMCJ) and Vidi grant (IP and LMCJ).
JFR and CPR gratefully acknowledge funding through acknowledge the European Research Council (ERC consolidator grant NANOPRS, project 617266).

\bibliography{apssamp}

\end{document}